\documentclass[lettersize,journal]{IEEEtran}
\usepackage{amsmath,amsfonts}
\usepackage{algorithmic}
\usepackage{algorithm}
\usepackage{array}
\usepackage[caption=false,font=normalsize,labelfont=sf,textfont=sf]{subfig}
\usepackage{textcomp}
\usepackage{stfloats}
\usepackage{url}
\usepackage{verbatim}
\usepackage{graphicx}
\usepackage{cite}
\usepackage{hyperref}
\begin{document}

\title{Quantum Algorithms applied to\\ Satellite Mission Planning for Earth Observation}

\author{
Serge~Rainjonneau\IEEEauthorrefmark{2},
Igor~Tokarev\IEEEauthorrefmark{1},
Sergei~Iudin\IEEEauthorrefmark{1},
Saaketh~Rayaprolu\IEEEauthorrefmark{1},
Karan~Pinto\IEEEauthorrefmark{1},
Daria~Lemtiuzhnikova\IEEEauthorrefmark{1},
Miras~Koblan\IEEEauthorrefmark{1},
Egor~Barashov\IEEEauthorrefmark{1},
Mohammad~Kordzanganeh\IEEEauthorrefmark{1},
Markus~Pflitsch\IEEEauthorrefmark{1},
and~Alexey~Melnikov\thanks{\IEEEauthorrefmark{4}Corresponding author, e-mail: ame@terraquantum.swiss
\begin{center}
\fbox{
\begin{minipage}{0.45\textwidth}
Please check the published version, which includes all the latest additions and corrections: IEEE J. Sel. Top. Appl. Earth Obs. Remote Sens. 16:7062-7075, 2023, DOI: \href{https://doi.org/10.1109/JSTARS.2023.3287154}{10.1109/JSTARS.2023.3287154}\\
© © 2023 IEEE. Personal use of this material is permitted. Permission from IEEE must be obtained for all other uses, in any current or future media, including reprinting/republishing this material for advertising or promotional purposes, creating new collective works, for resale or redistribution to servers or lists, or reuse of any copyrighted component of this work in other works.
\end{minipage}
}
\end{center}
}\IEEEauthorrefmark{1}\IEEEauthorrefmark{4}% <-this % stops a space
\\\IEEEauthorblockA{\IEEEauthorrefmark{2} Thales Alenia Space, 31100 Toulouse, France}
\\\IEEEauthorblockA{\IEEEauthorrefmark{1} Terra Quantum AG, 9000 St.~Gallen, Switzerland}
}

\maketitle

\begin{abstract}
Earth imaging satellites are a crucial part of our everyday lives that enable global tracking of industrial activities. Use cases span many applications, from weather forecasting to digital maps, carbon footprint tracking, and vegetation monitoring. However, there are limitations; satellites are difficult to manufacture, expensive to maintain, and tricky to launch into orbit. Therefore, satellites must be employed efficiently. This poses a challenge known as the satellite mission planning problem, which could be computationally prohibitive to solve on large scales. However, close-to-optimal algorithms, such as greedy reinforcement learning and optimization algorithms, can often provide satisfactory resolutions. This paper introduces a set of quantum algorithms to solve the mission planning problem and demonstrate an advantage over the classical algorithms implemented thus far. The problem is formulated as maximizing the number of high-priority tasks completed on real datasets containing thousands of tasks and multiple satellites. This work demonstrates that through solution-chaining and clustering, optimization and machine learning algorithms offer the greatest potential for optimal solutions. This paper notably illustrates that a hybridized quantum-enhanced reinforcement learning agent can achieve a completion percentage of 98.5\% over high-priority tasks, significantly improving over the baseline greedy methods with a completion rate of 75.8\%. The results presented in this work pave the way to quantum-enabled solutions in the space industry and, more generally, future mission planning problems across industries.
\end{abstract}


\section{Introduction}

The reliable functioning of Earth-orbiting satellites crucially affects our everyday services such as connectivity~\cite{chi_sur_2009}, navigation~\cite{dzi_sat_2007}, and media~\cite{mil_sat_2020}.  Most satellites receive dynamic instructions on executing their mission in orbit, and planning the exact sequence of tasks is critical to the efficiency and sustainability of the project~\cite{planning_and_replanning}. Algorithmic optimization solutions have been suggested~\cite{classical1,classical2,classical3} as a remedy to the planning problem. With the rise of quantum technologies, there is a need to explore how quantum computing can improve the time complexity or the quality of these solutions. This work focuses on planning the mission of Earth-orbiting imaging satellites using quantum machine learning~\cite{biamonte2017quantum,dunjko2018machine,qml_review_2023} and optimization. Specifically, it explores using near-term quantum technologies to improve the solution's effectiveness today. Similar approaches, such as the contribution of Ref.~\cite{annealing_sat}, suggested an algorithm to overcome the scheduling task using quantum annealers. Ref.~\cite{d_wave_paper} reviewed the literature and found that although innovative developments existed, none revealed any practical advantage over classical approaches. The current work explores the interplay between classical and gate-based quantum computing algorithms. The practical advantage of employing this hybrid approach was shown in our earlier contributions~\cite{asel_1,asel_2,asel_3}.

In general, space mission planning can be a computationally hard problem~\cite{planning_and_replanning} to solve; in large-scale missions, the size of the problem requires a prohibitive amount of computational resources. Finding an efficient plan requires the optimization of movement and the re-ordering of tasks to maximize the number of total completed tasks. In this work, each task is a request made to the satellite to capture an image of the surface of the Earth, and the aim is to maximize the number of images taken, given a list of all requests and a total available time. In this work, the imaging satellites orbit the Earth exactly on the Earth's terminator, which is the line separating Earth's sun-lit areas from the dark ones. In 24 hours, the satellite orbits approximately 15 times around the Earth, which leads to 15 orbits shown schematically in Fig.~\ref{fig:greedy_solution}(a). To capture the image, the satellite must continuously aim the camera at the target area for a time known as the acquisition slot. Each requested image has an allocated data-take opportunity (DTO) window. The satellite must point in the appropriate direction within this window for the entire acquisition period to accomplish a request. To aim at an area on Earth, the satellite needs to rotate its camera to point in the desired direction. The latter movement introduces a time delay that, when added to the acquisition time, could limit the overall agility of the satellite in covering multiple areas. The efficient satellite mission plan will be able to choose the order of the acquisition requests to maximize the number of completed requests.

This paper discusses the benefits of the optimization and reinforcement learning algorithms over a greedy baseline algorithm, the unique advantage offered by the potential of quantum computing, and a demonstration of how quantum methods can be applied to improve machine learning and optimization models.  Specifically, the novel quantum reinforcement learning approach shown in Sec.~\ref{sec:alphazero} offers a completion rate of 98.5\% on highest-priority requests in a multi-satellite system. The model hybridizes the AlphaZero approach in Ref.~\cite{alpha_zero} and is trained on the QMware quantum cloud~\cite{benchmarking}. Sec.~\ref{sec:satellite_mission_planning} introduces the practical problem setup in more detail, including the details on the data formatting in Sec.~\ref{sec:data_formatting}, re-ordering the requests in Sec.~\ref{sec:request_reordering}, relaying algorithm in Sec.~\ref{sec:relaying_algorithm}, and on solution chaining in Sec.~\ref{sec:solution_chaining}. Sec.~\ref{sec:results} offers three algorithms and their respective results: Sec.~\ref{sec:greedy_algo} establishes a baseline greedy algorithm, and Sec.~\ref{sec:optimization} and~\ref{sec:RL} discuss the optimization and reinforcement learning algorithms.  Finally, Sec.~\ref{sec:discussion} summarises the results.

\section{Satellite mission planning}\label{sec:satellite_mission_planning}

Each satellite's orbit is constrained to the Earth's terminator. Each satellite can only rotate at a maximum of one degree per second to orient itself appropriately to capture images within the acquisition window. Furthermore, the DTO duration for each area to be captured is defined by the arc ranging within 45$^{\circ}$ (referred to as the depointing angle) from the apex point, which is the point directly above the center of the request on Earth. As a result of the width of these arcs, multiple requests can have substantial overlap in their acquisition windows. Furthermore, while coordinates for image requests are provided in latitude and longitude, the satellite coordinates support an Earth-centric inertial (ECI) format, providing more spatial location support. Finally, two datasets are tested in this work, including a single-satellite set containing 462 requests and a two-satellite system with 2000 requests.  The main results of this work focus on the performance of various algorithms on the latter dataset and only for the requests with the highest priority.

\begin{figure}
\centering
\includegraphics[width=1\linewidth]{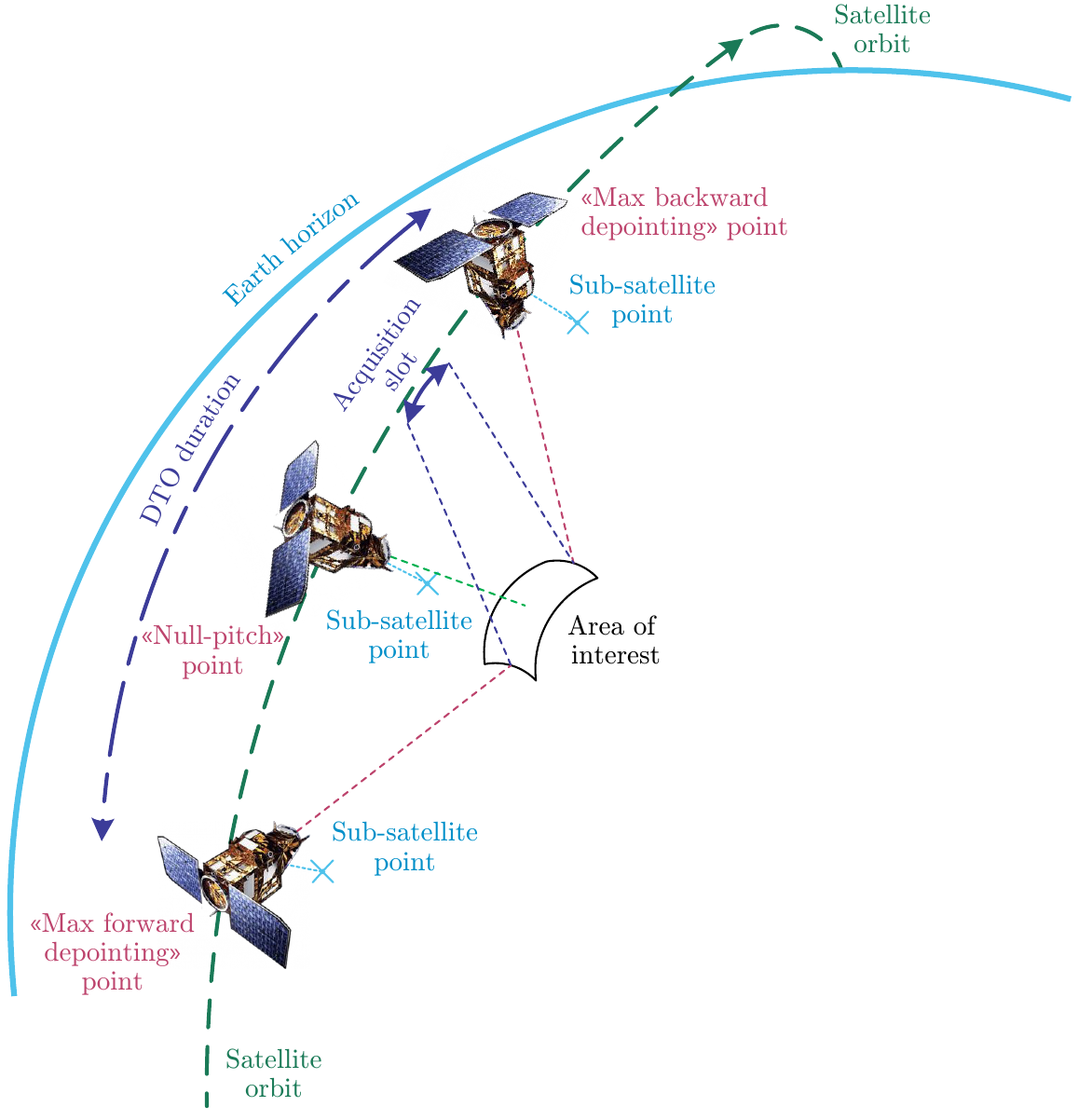}
\caption{The trajectory of a satellite orbiting on the Earth's terminator and across a DTO window. At the first and third positions on the orbital line, the satellite is at the ends of the request's DTO window, as its depointing angle is maximized at 45$^{\circ}$. At its second position, it is at the apex point, directly above the request location.}
\label{Fig1}
\end{figure}

\subsection{Data Formatting}\label{sec:data_formatting}

Two information sets were used in the preparation of this paper:

\begin{itemize}
    \item information about the satellite motion, including:
    \begin{itemize}
        \item the orbit number,
        \item time stamp, and
        \item satellite position and velocity in Cartesian ECI coordinates; and
    \end{itemize}
    \item information about acquisition request, including:
    \begin{itemize}
        \item request ID,
        \item request priority ranging from 1 to 4, where 4 denotes the lowest priority,
        \item start and end times of the DTO window of the request,
        \item the coordinates of the start and end of the median line,
        \item satellite ID, and
        \item Boolean values indicating the progress of the acquisition. 
    \end{itemize}
\end{itemize}

Fig.~\ref{Fig1} demonstrates the satellite movements during acquisition and the data that must be tracked during the capture, such as the acquisition angles, the points at which the DTO begins and ends, and the median coordinates.

\begin{figure*}
\centering
\includegraphics[width=1\linewidth]{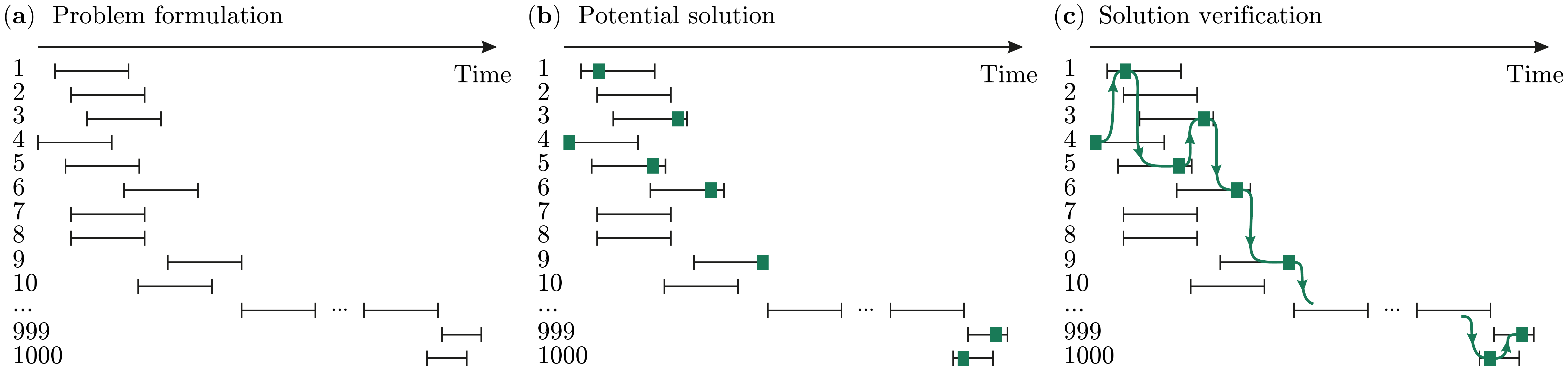}
\caption{(a) Visualization of DTO windows (horizontal bars) over a horizontal axis of time and a vertical axis indicating request index. (b) Visualization of solutions (green rectangles). (c) The chaining of requests based on their DTO windows and completion windows. As is evident, chaining provides a roadmap to connect all completed requests and map out the order of requests.}
\label{Fig2}
\end{figure*}

\subsection{Request Priority Ordering}\label{sec:request_reordering}

Considering the priority associated with each request, it is possible to express this objective as maximizing the completed requests in the order of priority; the highest priority requests are considered first, and only the lower priority requests are considered upon completion. The exact quantification compares the number of higher-priority requests accomplished. It moves on to the next priority in case of equality, but this work judges the algorithms by their completion rate performance purely on the high-priority $\pi_1$ request.

\subsection{Relaying Algorithm} \label{sec:relaying_algorithm}

The satellite must rotate according to the relaying algorithm to move from one request to another. To compute this rotation, two points of interest are the final median points of the first request and the initial median point of the second request. These points, respectively, signify the time at which the first request was completed and the time at which the next request will begin.  For a given transition, the relaying algorithm operates in two steps: 1) it computes the relative positions of each median point for the satellite, and 2) calculates in degrees the angle between these vectors.  Assuming that the satellite rotates at a constant speed of one degree every second and that the Earth is a perfect sphere, the resultant angle can be approximated as the total relaying time in seconds.

\subsection{Solution Chaining}\label{sec:solution_chaining}

Solution chaining considers the DTO windows, overlaps, and the windows' length compared to an acquisition time estimate. DTO windows and possible solutions within those windows are then visualized over an axis of time - see Fig.~\ref{Fig2}b. The most efficient preliminary method for this setting is to sort the requests by the start of their DTO windows and map the acquisitions. 

Next, the individual requests \textit{chained}: the intermittent time between captures is calculated using the relaying algorithm, after which the algorithm ensures sufficient time for the relaying movement between the completion of each pair of consecutive requests, ensuring that a sequence of requests can be executed in the provided time. Chaining enables the suggested solutions within the DTO windows of each request, which are otherwise independently scattered, to be connected together into one comprehensive solution for satellite movement. Shown in Fig.~\ref{Fig2}(c) is a visualization of the chaining process between multiple requests over time.

\section{Results: quantum algorithms for satellite mission planning}\label{sec:results}

% Fig.~\ref{fig3-algorithms} summarises the algorithms presented in the paper.
The table below summarises the algorithms considered in this work and their corresponding sections. 
\begin{figure}[h!]
    \centering
    \includegraphics[width=0.75\linewidth]{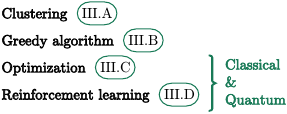}
    % \caption{Algorithms that are presented in this paper.}
    \label{fig3-algorithms}
\end{figure}
\begin{figure*}
\centering
\includegraphics[width=1\linewidth]{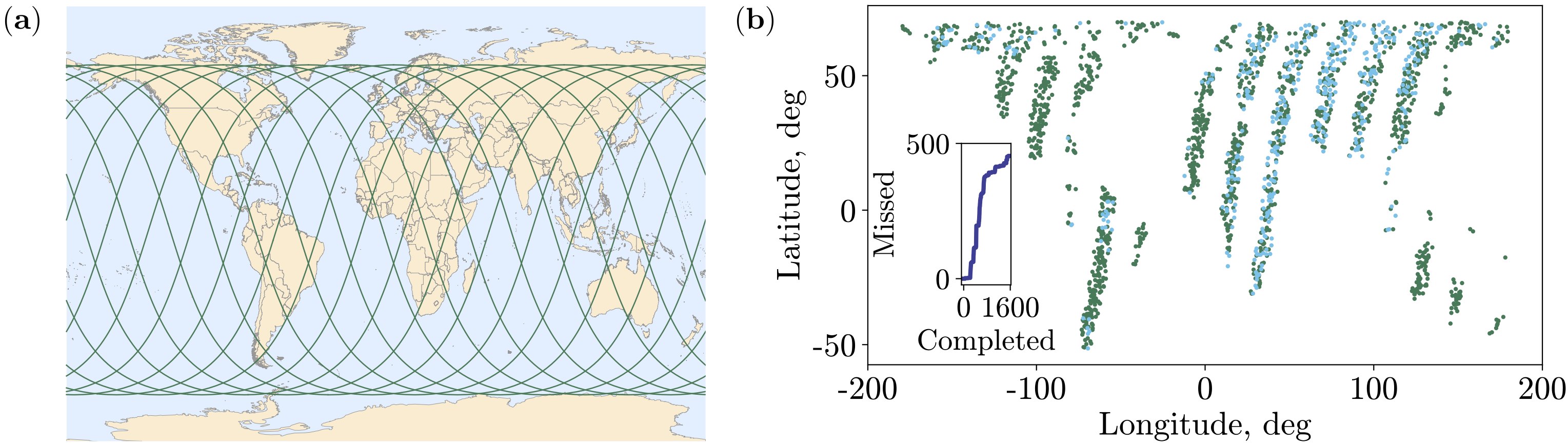}
\caption{(a) Schematic view of 15 satellite orbits performed within approximately 24 hours. (b) The results of the greedy algorithm are shown on the map. The task contains 2000 requests that are planned for acquisition using 2 satellites. Green points correspond to completed requests, whereas blue points are missed requests. (Inset) Progression of completed compared to missed requests.}
\label{fig:greedy_solution}
\end{figure*}

\subsection{Clustering for Pre-processing}
\label{sec:k-means}
To deal with increasing complexity in problems is to reduce the size of the dataset and deal with fewer parameters or data points at once. Therefore, clustering is useful to stratify the data by similarities and shorten the calculations necessary for the overall program. The simplest method to cluster the dataset is sorting the samples based on a single feature of the data, a method known as bunching. The bunching algorithms are explained in App.~\ref{sec:bunching}, but this section focuses on unsupervised-learning clustering methods such as the K-means algorithm~\cite{kmeans1,kmeans2}.

This algorithm is a well-known clustering algorithm, and its usage of physical distances in creating clusters makes it a natural fit for the Space Mission Planning problem. This algorithm is iterative, meaning it runs multiple iterations of the same steps and finally converges at a solution. The first step of this algorithm is the assignment stage; once $k$ random points are initialized to represent the $k$ cluster \textit{centroids}, each point in the dataset is assigned to the closest centroid. Once this is complete, the values of all points assigned to each centroid $k_n$ are averaged, and the value of $k_n$ is updated to the newly obtained mean value. 
Once these two steps are completed, they are repeated until the centroids no longer shift between two iterations (generally to an error threshold), which is when the final clustering for the dataset is attained.  The K-means algorithm clustered the dataset based on all the features of the requests: the DTO start and end times and the coordinates of the start and end of the median line.  This ensured that clusters bunch the requests based on their DTO windows and geographical locations.  The latter is crucial in efficiently using the relaying algorithm, as the geographically-close requests are simpler to inter-navigate.

\subsection{Greedy algorithm}\label{sec:greedy_algo}

The classical greedy algorithm is used to provide a reference solution. First, the orbital and request data are separated by satellite, and then the requests are clustered. In each cluster, the algorithm begins at the satellite's given start time and increments by one second until it enters the DTO window of the first request. Once the time stamp enters a DTO window, the algorithm checks for enough time to complete a request. If so, it completes the request, increments the time by the sum of the relaying time and the acquisition time, adds the request ID to a list of completed IDs and moves on to the next request. The algorithm chooses the highest priority request if multiple requests are available at a timestamp. If insufficient time to complete a request, it simply moves on to the next request and repeats the process. Once the final request is completed, or the timestamp exceeds the final DTO window, the algorithm outputs the percentage of completed requests of each priority.

An important limitation of the greedy algorithm is its bandwidth for anticipation, but it provides a baseline for comparison. Its simplicity allows it to run faster than other models, making it a good benchmark for data preprocessing. The results in Fig.~\ref{fig:greedy_solution}(b) and Tab.~\ref{tab:greedy} show the greedy algorithm struggles as the complexity of the input data increases. Therefore, to solve this problem, algorithms that can handle high levels of complexity without compromising accuracy are required.  The results of the greedy algorithm serve as the baseline algorithm in this work.  As evident in Tab.~\ref{tab:greedy}, the greedy algorithm solves the 1sat/462req dataset well but struggles on the 2sat/2000req dataset.  Thus, the latter was chosen as the comparison dataset as it allowed more possibilities for improvement.  Furthermore, it should be noted that only the top priority $\pi_1$ requests were taken as the baseline, so the models' accuracy performance will be compared with the baseline $\pi_1$ 2sat/2000req of 75.8\%.

\begin{table}
\caption{Results for the greedy algorithm by priority, run over two separate datasets.}
\label{tab:greedy}
\begin{tabular}{|l|l|l|}
\hline
\textbf{Priority} & \textbf{1sat/462req} & \textbf{2sat/2000req} \\ \hline
$\pi_1$           & 99.2\%            & 75.8\%                \\ \hline
$\pi_2$           & 88.6\%            & 36.1\%                \\ \hline
$\pi_3$           & 80.7\%            & 19.2\%                \\ \hline
$\pi_4$           & 74.0\%            & 18.2\%                \\ \hline
\end{tabular}
\end{table}

\begin{table}
\caption{Results of the greedy algorithm using classical K-Means clustering.}
\begin{tabular}{c|c|c|}
\cline{2-3}
&\multicolumn{2}{c|}{KMeans + Greedy} \\ \hline
\multicolumn{1}{|c|}{\textbf{Priority}} & \multicolumn{1}{c|}{\textbf{1sat/462}} & \textbf{2sat/2000req}\\ \hline
\multicolumn{1}{|c|}{$\pi_1$} & \multicolumn{1}{c|}{99.1\%} & 63.6\%  \\
\hline
\end{tabular}
\end{table}
    
\subsection{Optimization Methods} \label{sec:optimization}

As discussed earlier, one potential way to improve the runtime of an algorithm was to break the data into smaller clusters to reduce the necessary processing power. However, attempting to build a fundamentally more powerful algorithm, such as an optimization model, could be more fruitful. Optimization problems involve information and formulations, including graphs, movement patterns and permutations, and multiple viable solutions, out of which one ideal solution must be determined by the metrics and constraints of the problem. As such, optimization methods are well suited for mission planning problems akin to the one at hand, which aims to map out the best possible course of action for a system of satellites seeking to maximize the number of completed requests. Optimization, however, is expensive in terms of time and computation; consequently, the potential of harnessing the quantum advantage for optimization problems could be especially valuable in discovering and delivering a large speedup for mission planning problems.  

Some research for optimization for space mission planning currently exists. For instance, in Ref.~\cite{sez_opt_2019}, the problem of Earth observation from a satellite (EOS) is investigated, focusing on obtaining images of certain areas of the Earth's surface related to customer requests. An optimization approach for EOS's daily photo selection (DPSP) is proposed. DPSP is related to operational management and planning processes, where each photo the client orders brings profit. Still, not all requests can be satisfied due to physical and technological limitations. The objective is, therefore, to select a subset of queries for which the profit is maximized, and the proposed algorithm is based on the metaheuristic ant colony optimization algorithm (ACO). Examples based on real data are used as reference problems. The calculations show that the proposed algorithm can generate competitive and promising solutions.

The most well-known quantum-friendly optimization method is the quadratic unconstrained binary optimization (QUBO) model, which encapsulates the set of all optimization problems with the following attributes:
\begin{itemize}
    \item all variables are treated as binary objects
    \item constraints, while not necessarily absolute, are enforced accordingly by use of the reward function
\end{itemize}

This work explores the QUBO-adaptable formulation known as the \textit{integer optimization model} for the space mission planning problem. This solution has benefits and drawbacks but can serve as a viable solution to the satellite optimization problem. 

The integer optimization model is conceptualized by encoding the desirable outcome of the solution in the cost function, which should be maximized or minimized on the lattice of integer points of the special feasible subset in the multidimensional space. This feasible subspace is defined by adding the constraints of the satellite mission planning problem. One of the most powerful approaches to that problem is the \textit{branch and cut} algorithm \cite{land1960automatic}, used in solvers. In addition, the following functions are incorporated to streamline the calculations for the algorithm to stay within the constraints. Incorporating these functions is crucial to developing viable solutions as these functions, built over the Orekit space flight dynamics library~\cite{orekit}, are the key to accurate space mission planning.

\begin{itemize}
    \item \textit{Attitude Pointing:} given the satellite, timestamp, and a pair of longitude and latitude coordinates, this function returns the attitude (roll, pitch, and yaw) angles of the satellite when it is pointed towards the provided coordinates.
    \item \textit{Acquisition Duration:} given the starting and ending median coordinates of any request, this function returns the amount of time, in milliseconds, that the satellite would take to acquire the image.
    \item \textit{Maneuver Duration:} given the beginning and ending attitude angles (roll, pitch, and yaw), this function returns the amount of time necessary for the satellite to complete the maneuver from one attitude to the other.
    \item \textit{Read Ephemeris: } given a JSON file and a satellite ID, this function returns the satellite orbital information (directional speed, directional position, timestamps, etc.) in the form of an Orekit Ephemeris object, which is then used to perform calculations for other functions using the Orekit library.
\end{itemize}

\begin{figure*}
\centering
\includegraphics[width=1\linewidth]{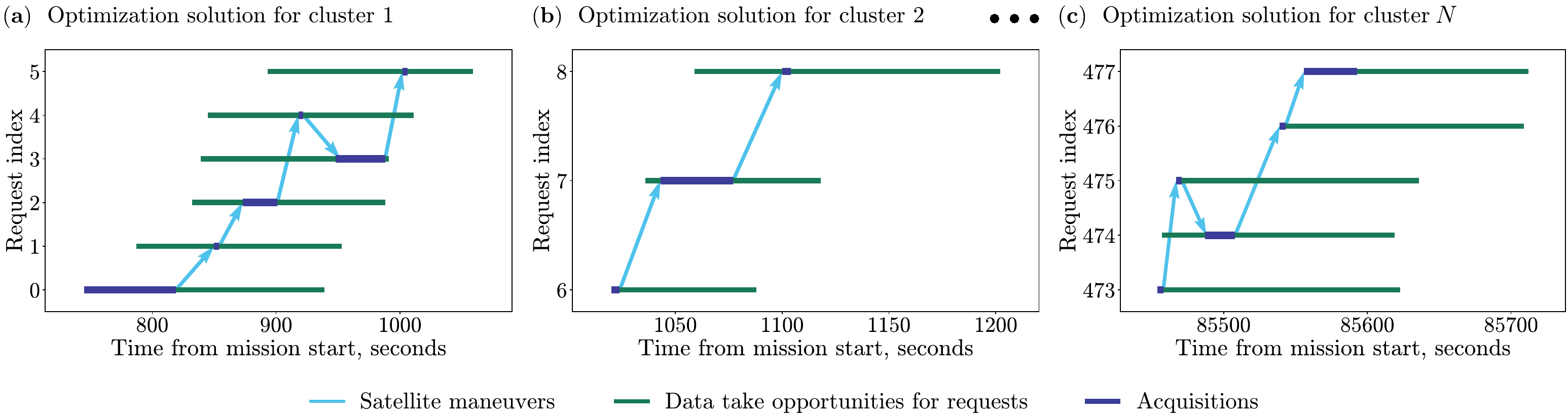}
\caption{Chaining results of classical integer optimization algorithm within clusters of data points.}
\label{optimization_chain}
\end{figure*}

Upon incorporating the above functions, revision of the constraints on this integer-based model, and modification of the clustering algorithm to generate a greater number of smaller clusters, the algorithm's runtime was maintained despite the significant increase in complexity. Given a set of requests, the variables central to this model include the priority of each request, the acquisition duration of each request, the start and the end of the DTO windows, and the location of the median points of the requests. From this, the algorithm calculates for each reasonable pair of the requests inside the small cluster all possible start times for the 1st request in a way that allows the relaying maneuver to the 2d request. This is achieved through simple iteration, using the Orekit functions described above: an initial value for the relaying maneuver duration is set to one second. Then, the maneuver start and end angles are calculated with the attitude pointing function. Afterwards, suppose the rotation time between the obtained angles via the maneuver duration function is not equal to or greater than one second, and the DTO limitations are not violated. In that case, meaning the relaying time is equal to one second. Otherwise, the duration is increased by one second, and this procedure is repeated until either the duration is appropriate or the DTO is violated. The result of these computations, $t_{min}$, is treated as the minimum relaying maneuver time.

Given a set of requests $f \in \mathbf{F}$, the priority of each request is denoted as $\pi_f$, and the acquisition duration as $\tau_f$. The start of the DTO window (release time) for request $f$ is represented as $r_f$ and the deadline of the DTO window is represented as $d_f$. In addition, $\mathbf{Q} \subset \mathbb{N}$ indicates the order of accepted requests. As a result, index $q \in \mathbf{Q}=\{0,\ldots, Q-1\}$ demonstrates the position in the queue of requests in which request $f$ is accepted. The discretized rotation of the satellite at the inception of the acquisition of request $f$ is indicated by $\alpha_{f}^{\text{start}} \in \mathfrak{A}_{f}^{\text{start}}$. Similarly, $\alpha_{f}^{\text{end}} \in \mathfrak{A}_{f}^{\text{end}}$ is the discretized rotation of the satellite after the acquisition of the request. However, all possible inceptions $b_{f\alpha}$ for requests can be streamlined in order of increasing time. Note that $b_{f\alpha}$ is the set of all possible points in time at which it is possible to complete the $f_1$ request and then move to the $f_2$ request. That is, it is at a time greater than $r_f$. Thus all the possible angles will correspond to these moments. The set of possible pairs is defined as $\mathbf{L}$, such that the maneuver $f_1 \to f_2$ is possible; i.e.:
\begin{equation}
    \exists b_{f_1\alpha_{f_1}}, b_{f_2\alpha_{f_2}}: b_{f_1\alpha_{f_1}} + t_{min}(b_{f_1\alpha_{f_1}} + \tau_{f_1}, f_1, f_2) + \tau_{f_1} = b_{f_2\alpha_{f_2}},
\end{equation}
where $t_{min}(t, f_1, f_2)$ is the minimum amount of time required for a satellite to rotate from its position at the end of $f_1$ at the moment $t$ to the starting point of $f_2$. For all $(f_1, f_2) \in \mathbf{L}$, the sets $\mathfrak{B}_{f_1 f_2}^{start} \subseteq \mathfrak{A}_{f_1}^{end}$ and $\mathfrak{B}_{f_1 f_2}^{end} \subseteq \mathfrak{A}_{f_2}^{start}$ and mapping $M_{f_1, f_2}: \mathfrak{B}_{f_1 f_2}^{start} \to \mathfrak{B}_{f_1 f_2}^{end}$ are defined such that $M_{f_1, f_2}(\alpha_1) = \alpha_2$, if $b_{f_1\alpha_{1}} + t_{min}(b_{f_1\alpha_1} + \tau_{f_1}, f_1, f_2) + \tau_{f_1} = b_{f_2\alpha_{2}}.$

The binary variable $x_{fq}$ is introduced such that:

\begin{equation}
    x_{fq}=\begin{cases}
            1, & \text{if $f$-th request is  started  in the $q$-th slot,}\\
            0, & \text{otherwise.}
    
            \end{cases}
\end{equation}

In addition, the variable $y_{f\alpha}$ is also introduced, where:

\begin{equation}
    y_{f\alpha}=\begin{cases}
            1, & \text{if angle $\alpha$ is the start angle for request $f$},\\
            0, & \text{otherwise.}
    
            \end{cases}
\end{equation}

Finally, the indicator variable $\kappa_{f_1 f_2}$ shows if requests were completed successively one after another:

\begin{equation}
    \kappa_{f_1 f_2} = 
        \begin{cases}
        1, & \text{if $f_2$-th  follows right after the $f_1$-th,}\\
        0, & \text{else}.
        \end{cases}
\end{equation}

The cost function is defined as:
\begin{equation}
    \sum\limits_{f\in \mathbf{F}}\left( \sum\limits_{q\in \mathbf{Q}} J_{f}x_{fq} - \sum_{\alpha \in \mathfrak{A}_{f}^{start}} \gamma_{f\alpha} y_{f\alpha}\right)
    \to \max,
\end{equation}
where $J_{f}$ is the weight of request $f$, and $\gamma_{f\alpha}$ is a coefficient needed for penalizing time consuming solutions. $J_f = 1$ was selected for the lowest priority requests in each cluster. For the higher priority requests, the weight is greater than the sum of all lower priority requests' weights by one since each higher priority request is valued more than any amount of lower priority requests. The coefficients $\gamma_{f\alpha}$ range from $0$ to $\frac{1}{Q}$ for the following definition:
\begin{equation}
   \gamma_{f\alpha} = \frac{b_{f\alpha} - r_f}{Q(d_f-r_f-\tau_f + 1)}.
\end{equation}
It guarantees that $\sum_{f}\sum_{\alpha \in \mathfrak{A}_{f}^{start}} \gamma_{f\alpha} y_{f\alpha} < 1$, and consequently, the completion of the lowest priority request will be more important than the particular order of requests, but the earliest possible completion of each request is preferable.

As was the case with the other optimization model, more than one request cannot be executed in the same order, and each request should be completed not more than once:
\begin{equation}
    \forall f \in \mathbf{F}, \sum\limits_{q\in \mathbf{Q}} x_{fq} \leq 1 
\end{equation}
\begin{equation}
    \forall q \in \mathbf{Q}, \sum\limits_{f\in \mathbf{F}} x_{fq} \leq 1 
\end{equation}

Any completed request is started with one particular possible angle:

\begin{equation}
        \sum\limits_{\alpha\in \mathfrak{A}^\mathrm{start}_f} y_{f\alpha} = \sum\limits_{q\in \mathbf{Q}} x_{fq} \quad \forall f \in \mathbf{F}.
    \end{equation}

All requests are completed one after another without empty slots in line until the satellite stops:
    \begin{equation}
        \sum\limits_{f_1\in \mathbf{F}} x_{f_1(q-1)} \geq \sum\limits_{f_2\in \mathbf{F}} x_{f_2q} \quad \forall q > 0.
    \end{equation}

To evaluate the variable $\kappa_{f1 f_2}$ appropriately, the following system of linear equations is introduced, and each excludes the impossible values of $\kappa_{f_1 f_2}$. Firstly, if requests $f_1$ and $f_2$ were completed straight one after another, then $\kappa_{f_1, f_2}=1$:
\begin{equation}
\kappa_{f_1 f_2} + 1 \geq x_{f_1, q-1} + x_{f_2, q} \quad \forall f_1, f_2 \in \mathbf{F}, q > 0.
\end{equation}

Each request is followed by not more than one request, and each request follows after not more than one request:

\begin{equation}
    \sum_{f_1 \in \mathbf{F}}\kappa_{f_1, f_2} \leq 1 \quad \forall f_2 \in \mathbf{F},
\end{equation}
\begin{equation}
    \sum_{f_2 \in \mathbf{F}}\kappa_{f_1, f_2} \leq 1 \quad \forall f_1 \in \mathbf{F}.
\end{equation}

If one request follows another one, then each of these requests was completed in some order:
\begin{equation}
    \kappa_{f_1 f_2} \leq \sum_{q > 0}x_{f_2, q} \quad \forall f_1, f_2 \in \mathbf{F},
\end{equation}
\begin{equation}
    \kappa_{f_1 f_2} \leq \sum_{q < Q-1}x_{f_1, q} \quad \forall f_1, f_2 \in \mathbf{F}.
\end{equation}

With the use of predefined mapping $M_{f_1, f_2}$, if the request $f_1$ starts with angle $\alpha_1$, and the request $f_2$ is the next one, then the next acquisition start angle is fixed:
\begin{equation}
    y_{f_2\alpha_2} + 1 \geq \kappa_{f_1, f_2} + y_{f_1\alpha_1} \quad \forall \alpha_1 \in \mathfrak{B}_{f_1 f_2}^{start} \text{,} 
\end{equation}
where $\alpha_2 = M_{f_1, f_2}(\alpha_1).$

Additionally, maneuver $f_1 \to f_2$ is possible only with particular initial angles for $f_1$. Otherwise, the satellite will not have enough time to finish the acquisition of the request $f_1$ and move to the $f_2$:

\begin{equation}
    \sum_{\alpha_1 \in \mathfrak{B}_{f_1 f_2}^{start}}y_{f_1\alpha_1} \geq \kappa_{f_1 f_2} \quad \forall (f_1, f_2) \in \mathbf{L}.
\end{equation}

On the other hand, if $f_1 \to f_2$ is an impossible transaction, then

\begin{equation}
    \kappa_{f_1, f_2} = 0 \quad \forall (f_1, f_2) \not \in \mathbf{L}.
\end{equation}

The final constraint fixes the acquisition start angle for the 1$^{st}$ request in a queue as the earliest possible angle:

\begin{equation}
    y_{f \alpha_0} \geq x_{f 0} \quad \forall f \in \mathbf{F},
\end{equation}
where $\alpha_0$ is an angle corresponding to the beginning of the DTO for request $f$.

In contrast to the greedy algorithm, which considered the requests in the order of open DTO windows, this model is more intelligent in finding the best path to fit the maximal number of requests. In the end, the $\pi_1$ requests reached 98.1\% completion using the Gurobi solver~\cite{gurobi}. The solution obtained through optimization is partly depicted in Fig.~\ref {optimization_chain}.  Furthermore, the clusters are connected, calculating the minimum relaying maneuver time from the last request of the previous cluster to that of the next. This is illustrated in Fig.~\ref {optimization_chain} as cutting the beginning of every DTO from the start if it proves impossible to rotate towards the request in this period. After this procedure, the next cluster can be treated as independent.

One of the greatest benefits of this model is its compatibility with both near-term and long-term quantum technology. As a linear optimization model that uses a grid of binary parameters, it can be transformed into QUBO, as it is shown in App.~\ref{sec:QUBO}, and fit rather quickly to the quantum Ising model, a model containing arrays of qubits in a grid where their spin states depend on their neighbors. Furthermore, as this model functions akin to a minimization problem, it would be extremely efficient to run on a quantum annealing machine \cite{annealing}, which could solve optimization problems by slightly changing the Hamiltonian from a given initial state with a known minimum to a new state, representing the optimal solution. However, it is complicated to use the satellite mission planning problem in the QUBO form via currently available classical or quantum devices. For example, both D-Wave's Leap Hybrid solver \cite{dwave} and Gurobi have difficulties solving even a small cluster with 4 requests over a time limit of 1 minute. Still, linear programming can obtain the solution for the 2000 requests and 2 satellites dataset in less than 3 minutes. This runtime was achieved by decomposing the clusters so that their final size was small enough to achieve approximately the same time as the greedy algorithm. Note that 
if the same number of requests were simultaneously considered, the runtime would be noticeably longer.

\begin{figure*}
    \centering
    \includegraphics[width=0.85\linewidth]{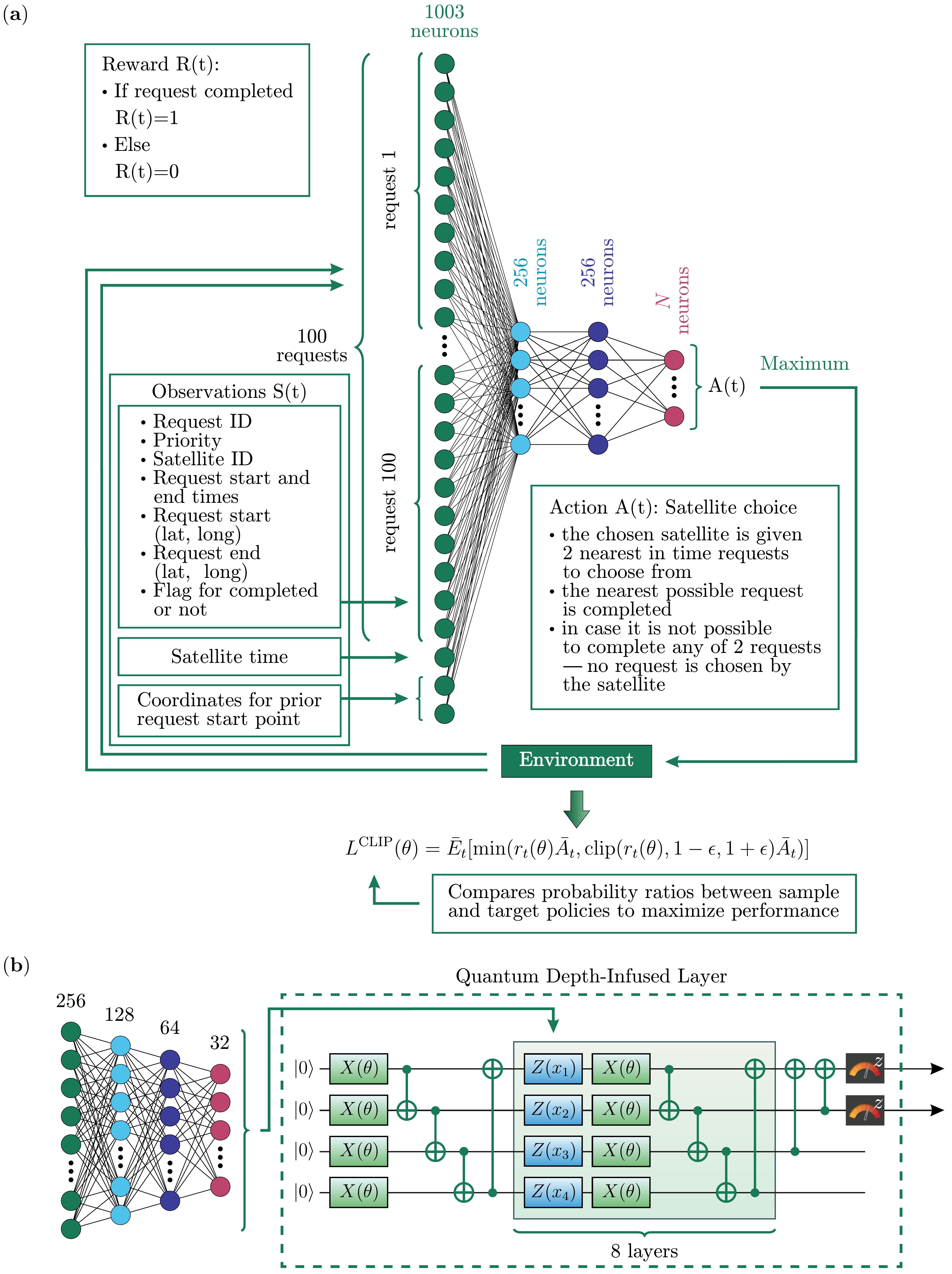}
    \caption{(a) The reinforcement learning PPO model used to solve the mission planning problem. The state of the model encompasses the data of 100 requests, which are fed into the MLP agent to output an action (one request that is selected to complete), which then feeds into the environment (clip equation), generates an appropriate reward, and updates the state. (b) Quantum-hybrid reinforcement learning model. A quantum circuit (left) is added to the beginning of the MLP Agent in the RL model (right) to incorporate quantum computation into the neural network.}
    \label{fig:ppo}
\end{figure*}

\subsection{Reinforcement Learning} \label{sec:RL}

Reinforcement learning (RL) is a machine learning paradigm in which an agent interacts with some environment and trains through informed trial and error. The RL training algorithm uses reward functions to assign value to the agent's actions in any state of the environment. Generally, a state can have constraints and features. The RL agent can use a policy model to decide on an action given a state, which subsequently affects the environment and transforms the state. Suppose the resultant state of the environment is engineered by the data scientist to contain a positive reward. In that case, the policy model is trained to take the appropriate action to maximize the probability of achieving that reward. The Environment is a function of a triplet of variables $(S, A, P)$, where $S$ is a state space, $A$ is an action space, and P is a transition function. When the reward function, $r$, is factored in, a Markov Decision Process (MDP) is generated with the property ($S$, $A$, $P$, $r$),  $r: S \times A \rightarrow \mathbb{R}.$ The Agent starts from state $s_0$ and takes action $a_0$, for which the reward $r_0$ is obtained in each step of training and subsequently trains by producing trajectories $T:= (s_0,a_0,r_0,s_1,a_1,r_1,s_2,a_2,r_2,\cdots).$

\begin{figure*}
    \centering
    \includegraphics[width=1\linewidth]{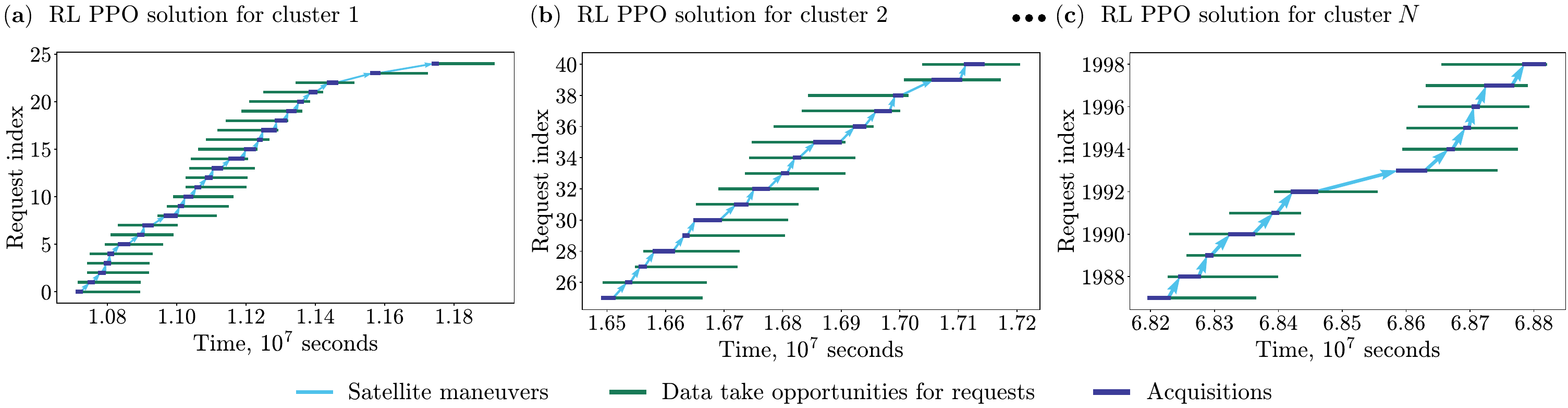}
    \caption{Visualization of the chaining process used to complete requests within each cluster of the PPO algorithm.}
    \label{rl_chain}
\end{figure*}

\subsubsection{RL Environments}

Two environments are developed for this AEOS task: satellite- and request-centred environments. For each environment, requests are sorted by their DTO open time. 

\textit{The satellite-centred environment:} the agent views the problem from the satellite's perspective and considers the 100 closest (by DTO open time) data points for each satellite, each utilizing 10 data parameters. Moreover, three additional features were also made available to the agent: the current time for the satellite and the latitude and longitude of the starting point of the last completed request, creating a total observation space of size 1003. The number of nearest requests is a variable that depends on task size. A Boolean flag is one of the features kept for each request, marking each data point as either complete or incomplete to avoid redundancy, and the observation space is made to only consider requests that are marked as incomplete to expedite the computations required by the agent, which is built as a neural network.

Once the request is selected, the agent is rewarded with 1 if that request is completed and 0 otherwise. With this process recurring, the agent attempts to complete as many requests as possible until the time of the satellite is greater than the DTO windows of all remaining requests, meaning they can no longer be completed. The agent works with each satellite and predicts which request must be done. The data used in this environment was artificially generated.
\textit{The request-centred environment:} at each step, the agent views the problem from the perspective of a request, deciding which satellite is best suited to complete it. The 5 nearest request options are determined by DTO open time for each satellite, and the request to execute is chosen by the minimum request execution time. This minimum time is calculated as the sum of the satellite timestamp, maneuver duration and the acquisition time due to solution chaining. It must be less than the request DTO end time for completion. This procedure is then iterated with the next batch of 5 nearest satellites.

\subsubsection{Proximal Policy Optimization}
The Proximal Policy Optimization (PPO) algorithm, shown in Fig.~\ref{fig:ppo} and first introduced in Ref.~\cite{ppo_paper}, was implemented to provide a mission planning policy. In RL, a policy is an operation that maps an action space to a state space. The agent learns the best action for each situation by calculating the policy gradients. In other words, the agent uses gradient descent to calculate the expected value of each action at a certain state space and determine which action has the likelihood of the highest reward. The equation for the PPO algorithm is as follows:
\begin{equation}
    L(\theta) = \bar{E}_t[\mathrm{min}(r_t(\theta)\bar{A}_t, \mathrm{clip}(r_t(\theta),1-\epsilon,1+\epsilon)\bar{A}_t)]
\end{equation}
where $\bar{E}_t$ is the current expected value of the policy, $\theta$ is the policy parameter, $\epsilon$ is the hyperparameter, $\bar{A}_t$ is the estimated advantage provided by the PPO at time $t$, and $r_t(\theta)$ is the importance sampling ratio. This ratio, derived from the Monte Carlo sampling methods~\cite{ppo_paper}, denotes the ratio of probabilities under both the old and the new policies. By keeping the value of $\epsilon$ small, the model ensures that the update on the policy at each increment is not too large; as a consequence, the learning done by the model stays relevant. 

\begin{figure*}
    \centering
    \includegraphics[width=0.9\linewidth]{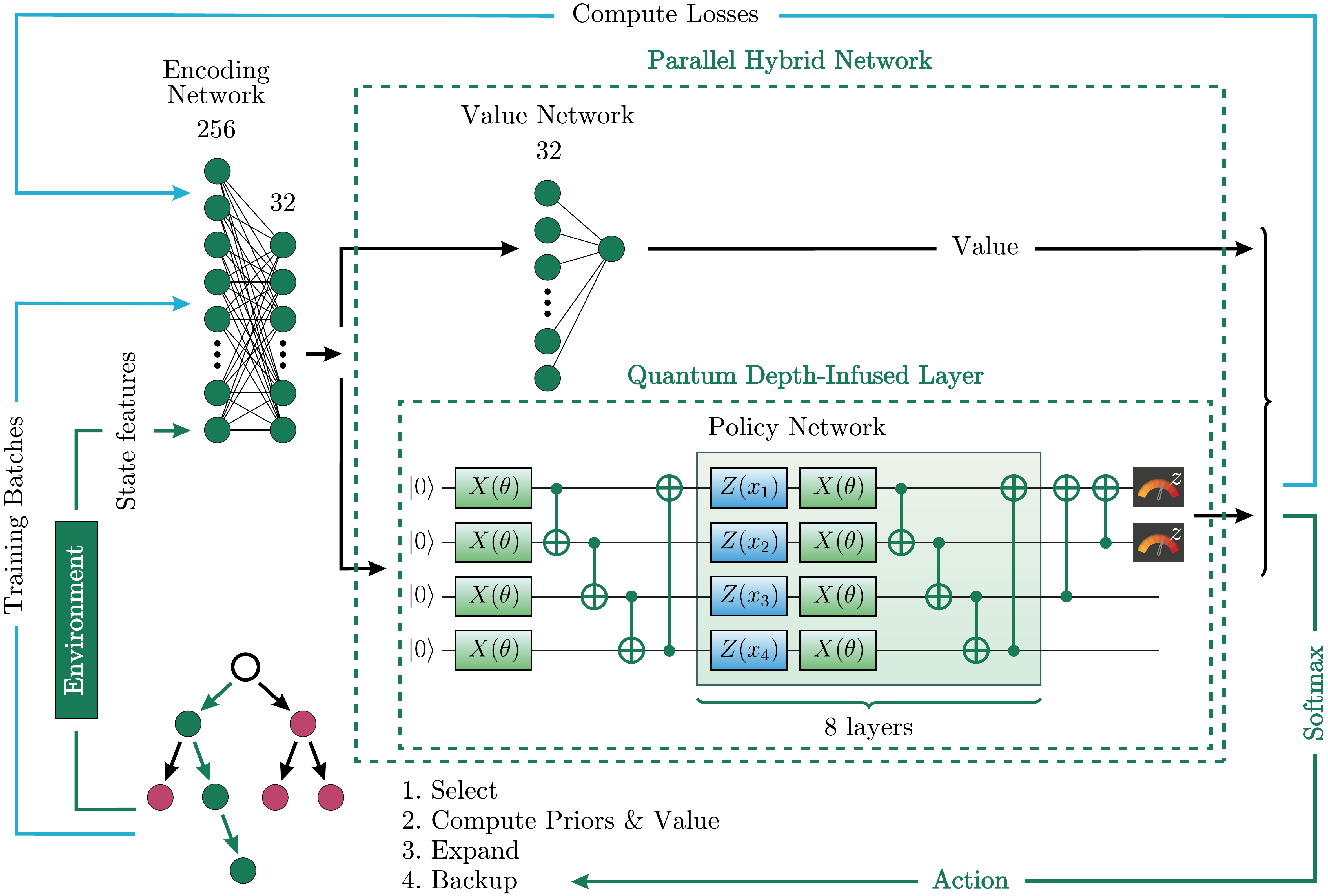}
    \caption{The hybrid AlphaZero architecture used to achieve a near-optimal solution to the satellite mission planning.}
    \label{fig:alphazeo}
\end{figure*}

\subsubsection{Hybrid-quantum PPO}\label{sec:hybrid_ppo}

Inspired by the quantum advantage shown in Refs.~\cite{jer_par_2021,sko_qua_2022,asel_1,asel_2,asel_3,senokosov2023quantum,sedykh2023quantum}, this section investigates the utility of a hybrid-quantum neural network as a policy model for RL. Fig.~\ref{fig:ppo}(b) illustrates the specific policy network within a hybrid MLP model. In the quantum-classical network, the output of the classical neural network is used as inputs to a parametrized quantum circuit (PQC). PQCs are quantum circuits that use parametrized quantum gates~\cite{benedetti,schuld2019quantum} such as Pauli rotations to encode data, $x$, and trainable parameters, $\theta$. The four-qubit PQC used in this work consists of three significant components: variational, data encoding, and measurement. The variational layer comprises a layer of Pauli-X rotations encoding four trainable parameters and ring-shaped CNOTs for entanglement. The data encoding layer embedded four features in parallel using Pauli-Z rotations, and the measurement layer comprised two Z-basis measurements on the first two qubits. In sequence, four qubits were initialized in the ground state, and then a variational layer with randomly initialized parameters was appended, followed by eight repetitions  of encoding and variational layers. Each of the eight encoding layers encoded four features of the dataset, which created a lattice of 32 features across the four qubits. Finally, the measurement layer was included to produce two classical real-valued outputs. Fig.~\ref{fig:rl-results}(a) shows this improvement in practice: the hybrid quantum neural network achieves a higher reward in only 8k steps which remains inaccessible to a classical network of the same complexity even after 110k environment steps. Notably, these two networks were trained five times with varying initialization points, and the plot in Fig Fig.~\ref{fig:rl-results}(a) only shows the best runs of the hybrid and the classical models. The reader might notice that the hybrid model starts at a higher mean reward than the classical, a behaviour observed elsewhere in the literature~\cite{asel_1}.  Additionally, the quick ascent aligns with the findings of Ref.~\cite{caro2021generalization}, suggesting that quantum models can generalize from a few data points. 

\subsubsection{Hybrid AlphaZero}\label{sec:alphazero}

The optimal results are expected from picking the best candidate from part of the solution pipeline. Specifically, in Sec.~\ref{sec:RL}, it was shown that reinforcement learning is a powerful algorithm that can be boosted in training and solution optimality through hybridization with quantum models.  This section shows hybridized AlphaZero~\cite{alpha_zero}. 

\begin{figure*}
    \centering
    \includegraphics[width=0.85\linewidth]{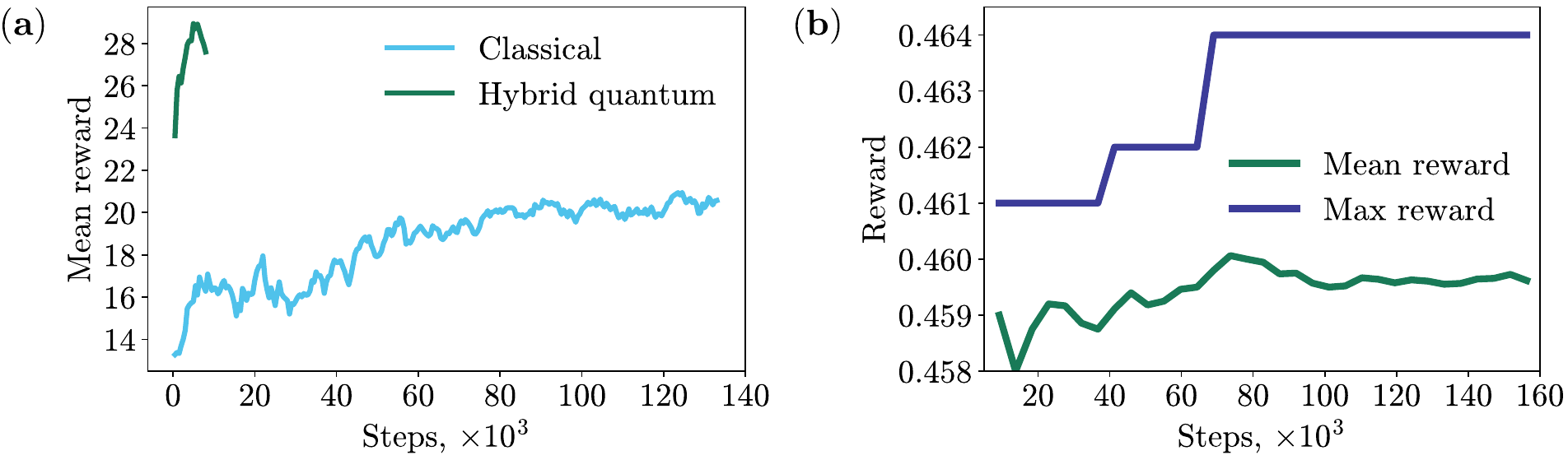}
    \caption{(a) The rewards the classical and hybrid PPO RL agents achieve. The hybrid model reaches a higher reward at a significantly improved learning speed. (b) The agent reaches a maximum normalized reward of 0.464 after 70,000 steps of the simulation. This is compared to a total possible reward of 0.471, translating to an accuracy of 98.5\%.}
    \label{fig:rl-results}
\end{figure*}

The classical AlphaZero was a reinforcement learning (RL) algorithm developed by DeepMind that showed promise in solving difficult RL problems, such as chess, shogi, and go~\cite{alpha_zero}. AlphaZero uses a computational tree of environment states whose values and probabilities are determined from the outputs of value and policy networks. The AlphaZero implementation in this work comprises four parts: a Monte Carlo tree search (MCTS), an encoding network, a policy network, and a value network. 

The foundation of this model is akin to a Monte Carlo Tree Search (MCTS) model, which is used primarily for path prediction problems and board games based on strategy~\cite{mc1,mc2}. The four major components of MCTS are selection, expansion, simulation, and backpropagation. The model begins at the tree's root and selects optimal child nodes until it reaches a leaf. Once it reaches the leaf, it expands the tree, creating a new child node. Then, the model simulates the remainder of the path-finding process from the newly created node and finally backpropagates with the newfound information to update the hyperparameters of the tree. The hybrid AlphaZero model uses MCTS with a parametrized quantum circuit as its policy network. The MLP agent of the model first recommends a state action. Then the action is applied to the environment, generating a reward and an updated state. However, in this case, the environment is the MCTS model; the recommended action is simulated in the model. Then, the backpropagation step of the MCTS is used to update the weights within the tree itself to refine its recommendations.

\begin{table*}
\caption{A summary of the performance of the algorithms on the 2000 requests task performed with 2 satellites. The comparison demonstrates the greedy algorithm as a baseline, the hybrid AlphaZero, and integer optimization models.  }
\label{tab:concluding_results}
\begin{tabular}{|c|c|c|c|c|c|}
\hline
Model               & Greedy only  & Greedy + KMeans  & \textbf{RL (Hybrid AlphaZero)} & Optimization \\ \hline
$\pi_1$ Completion \% &75.8\% & 63.6\%             & \textbf{98.5}\%          & 98.1\%       \\ \hline
\end{tabular}
\end{table*}

As each iteration of this loop occurs, there is also an outer loop in this algorithm. That outer loop is primarily concerned with the loss values being optimized; as a result, once the agent recommends an action, the outer loop takes training examples of a certain size and simulates the results over these batches to calculate the policy loss, value loss, and total loss. Once these values are calculated, the neural network weights are adjusted and fed back into the original loop, where the cycle repeats until the loss values are optimized. 

Fig.~\ref{fig:alphazeo} illustrates the inner workings of the hybrid AlphaZero model employed in this paper. The state variables are passed into an initial encoding network with 2 hidden layers of sizes 256 and 32 neurons. The information is then passed in parallel~\cite{kordzanganeh2023parallel} to 1) a single neuron using a fully-connected layer and 2) a policy network as a PQC. The PQC resembles the one explained and implemented in the PPO model in Sec.~\ref{sec:hybrid_ppo}. Fig~\ref{fig:rl-results}(b) shows the training performance of this model (best out of five tries), which achieves a completion rate of $\pi_1$ = 98.5\%. Presenting the highest $\pi_1$ completion rate on the 2000 requests and 2 satellites dataset of any other model explored in this paper. The completion rates of the highest-performing algorithms from each section on this dataset are displayed in Fig.~\ref{tab:concluding_results}.

\section{Discussion}\label{sec:discussion}

This work provided two classes of solutions to the scheduling problem of satellite mission planning: optimization and hybridized reinforcement learning. From each class, the best-performing candidates were the integer optimization model and the hybrid AlphaZero, respectively. The dataset with 2 satellites and 2000 was used as a performance benchmark for the algorithms presented in this work. The optimization and hybrid AlphaZero algorithms achieved $\pi_1$ completion rates of 98.1\% and 98.5\%, respectively, while the greedy algorithm only exhibited a 78.5\% 
 (63.6\% with k-means clustering) completion rate. In the single satellite model, the optimization algorithm reached 100\% completion on $\pi_1$, $\pi_2$, and $\pi_3$ requests and 96.2\% completion on $\pi_4$ requests in 6 minutes. This work showed that by using reinforcement learning and optimization models, it is possible to improve the results of mission planning that are otherwise obtained through simple greedy models. This work presents a step towards creating quantum-enhanced solutions in the space industry.

\bibliography{lib.bib}
\bibliographystyle{IEEEtran}

\clearpage
\appendix
\subsection{Bunching algorithms}\label{sec:bunching}
Below are a few bunching algorithms used to cluster the data purely by sorting with respect to certain dataset features.
\subsubsection{DTO Bunching}
This algorithm first sorts the data by the DTO start times. After that, the data points are clustered by DTO overlap. In other words, all entries in any given cluster share at least one portion of their DTO windows with $all$ other requests in the same cluster. As DTO windows are defined by the time when the satellite is in a position to capture the request, this method also functions as a form of geographical clustering due to the correlation of the location and DTO times.
\subsubsection{Priority Bunching}
In this algorithm, the requests are simply clustered by their priority ranks (1-4) and then ordered by the DTO start times. The utility of this lies greatly in the importance of the priority rankings, as the algorithm operates on the premise of each priority rank request being worth $n$ times a request of the next highest priority rank, where $n$ is the order of magnitude between two priority sets (for instance, a priority 1 request would be $n^2$ times as valuable as a priority 3 request).
\subsubsection{Bunch Sort}
This algorithm puts together the two clustering algorithms above, in a sense. First, the data is sorted by DTO windows, after which it is clustered by the DTO overlap. However, after that, the data is sorted within each cluster, this time by the end of their DTO windows and then by priority index. Consequently, data points with the same priority in the same cluster are ranked by the end of their DTO windows, maintaining the DTO structure within the priority indexing structure.
\subsection{QUBO formulation} \label{sec:QUBO}

Adiabatic quantum computers can approximately solve NP-hard problems, such as quadratic unconstrained binary optimization, faster than classical computers. Since many machine learning problems are also NP-hard, adiabatic quantum computers might be instrumental in training machine learning models efficiently in the post-Moore’s law era. To solve problems on adiabatic quantum computers, they must be formulated as QUBO problems, which is possible by several techniques. This paper formulated the problem as a QUBO problem, making it conducive to being trained on adiabatic quantum computers. Since all the constraints will go into the cost function with some coefficients, they must be algebraically reformulated. Below are examples of how these constraints change according to the QUBO formulation.

\begin{gather}
    \forall f \in \mathbf{F}, \sum\limits_{q\in \mathbf{Q}} x_{fq} \leq 1 \Rightarrow \left(\sum\limits_{q\in \mathbf{Q}} x_{fq} - \frac{1}{2}\right)^2,
\end{gather}

\begin{gather}
    \forall q \in \mathbf{Q}, \sum\limits_{f\in \mathbf{F}} x_{fq} \leq 1 \Rightarrow \left(\sum\limits_{f\in \mathbf{F}} x_{fq} - \frac{1}{2}\right)^2,
\end{gather}

\begin{gather}
    \forall f \in \mathbf{F}, \sum\limits_{\alpha \in \mathfrak{A}_f^{start}} y_{f\alpha} = \sum\limits_{q\in \mathbf{Q}} x_{fq} \Rightarrow \\ \left(\sum\limits_{\alpha \in \mathfrak{A}_f^{start}} y_{f\alpha} - \sum\limits_{q\in \mathbf{Q}} x_{fq}\right)^2,
\end{gather}

\begin{equation}
\begin{gathered}
    \forall q > 0, \sum\limits_{f_1 \in \mathbf{F}} x_{f_1 q-1} \geq \sum\limits_{f_2 \in \mathbf{F}} x_{f_2 q} \Rightarrow \\
    \sum\limits_{f \in \mathbf{F}} (x_{f_2 q} - x_{f_1 q-1}) + \sum\limits_{i} 2^i z_i = 0,
\end{gathered}
\end{equation}
where $z_i$ is a slack binary variable, and the number of slack variables depends on the problem. Thus, the idea of the QUBO formulation is to reduce all constraints to linear equalities, square them and then pose as penalties in the cost function. The more rigorous analysis of the slack variable implementation is provided in \cite{workflow}.
QUBO formulation is written in general terms as:
\begin{equation}
    Q = Q_0 +\beta C, \ C \geq 0,
\end{equation}
where $Q$ is a cost function, $Q_0$ is the objective function of the initial problem, $\beta$ are some coefficients that must be picked up, $C$ are non-negative definite quadratic constraints.
The solution then takes the following form:
\begin{equation}
    Q[q] = \min\limits_{\kappa} Q[\kappa].
\end{equation}
In the case where $C[q] > 0$, it is possible to choose such $\beta$, that if $q$ is unconstrained optimum, then it must be feasible.  Beta is chosen such that if $q$ is not feasible, then:
\begin{equation}
    Q[q] > Q[\text{feas}] = Q[0] = 0,
\end{equation}
where $Q[\text{feas}]$ means the initial zero solution.

The initial objective function that was used for the classical solution will take the following form for the quantum solution:
\begin{equation}
    Q_0 = - \sum\limits_{f\in\mathbf{F}}(\sum\limits_{q\in\mathbf{Q}} J_f x_{fq}^2 - \sum\limits_{\alpha\in\mathfrak{A}_f^{\text{start}}} \gamma_{f \alpha} y_{f \alpha}^2) \rightarrow \min.
\end{equation}
Also, the odds $\beta$ are chosen so that there is a better solution when the constraints are violated:
\begin{equation}
    C[q] > 0 \Rightarrow Q[q] \geq Q_0[q] + \beta > 0.
\end{equation}
Accordingly, in this case it is necessary to take $\beta > -Q_0[q].$ The boundaries of $q$ must be estimated, as the precise values are undetermined. The worst estimate was chosen as the following: 
\begin{equation}
    \beta > -Q[q] = -(\min\limits_{\kappa} Q_0[\kappa]) = -(-\sum\limits_{f} J_f) = \sum\limits_{f} J_f.
\end{equation}

\subsection{Greedy algorithm}
\begin{figure*}
    \centering
    \includegraphics[width=0.75\linewidth]{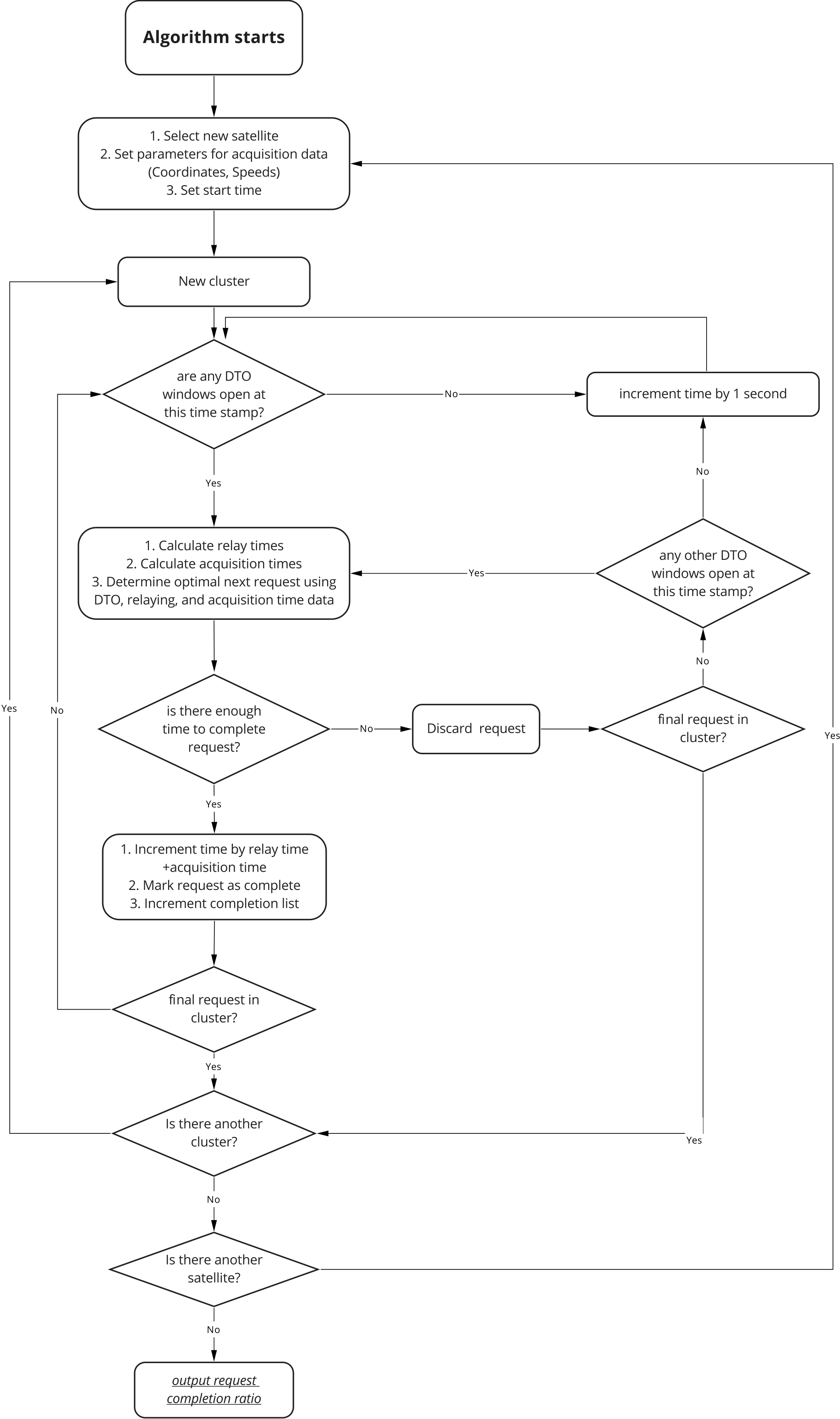}
    \caption{The flowchart used to implement the greedy algorithm in Sec.~\ref{sec:greedy_algo}. The time-incrementing block references the "open DTO windows" query. Without accessible DTOs within a cluster, satellites remain idle under the greedy algorithm, with the simulation advancing until availability arises. Upon addressing all requests, satellites transition to the subsequent cluster and replicate the procedure. Note that the connection between one-second time increments and new clusters is implicit rather than explicit.}
    \label{fig:greedy_algo}
\end{figure*}

Fig.~\ref{fig:greedy_algo} showcases the logic of the greedy algorithm from Sec.~\ref{sec:greedy_algo}. 

\vfill

\end{document}